# Hydrogen-doping mediated solid state thermal switch

R.J. Warzoha,[1,*] B.F. Donovan,[2] Y. Sun,[3] E. Cimpoiasu,[2] and S. Ramanathan[3]

[1]*Department of Mechanical Engineering, United States Naval Academy, Annapolis, MD, USA*

[2] *Department of Physics, United States Naval Academy, Annapolis, MD, USA*

[3] *School of Materials Engineering, Purdue University, West Lafayette, IN, USA*

**Recent reports reveal that isothermal chemical doping of hydrogen in correlated complex oxides such as perovskite nickelates (e.g. NdNiO$_3$) can induce a metal-to-insulator transition (MIT) without the need for temperature modulation. In this work, we interrogate the magnitude change in temperature dependence of thermal conductivity upon chemical doping of hydrogen, as any changes to the thermal properties upon doping offer a route to solid-state thermal switches as well as another potential signal to monitor in a diverse set of sensing, electronic, and optical applications. Using frequency-domain thermoreflectance, we demonstrate that a large concentration of hydrogen (~ 0.1 - 0.5 H/unit cell) completely suppresses the electronic contribution to thermal conductivity in NdNiO$_3$ thin films and reduces the phononic contribution by a factor of 2. These results are critical for the design of next-generation solid-state thermal switches, sensors, extreme environment electronics and neuromorphic memory architectures.**

Rare-earth nickelate perovskites (RNiO$_3$, R = rare-earth element) represent a class of complex correlated oxides that exhibit sharp metal-insulator transitions (MITs), around which electronic resistance can be regulated across several orders of magnitude [1–3]. Such materials are therefore candidates for next-generation electronic devices (e.g., neuromorphic computing [4] and memory devices [5], as well as environmental sensing [6].

For many applications it is often desirable to modulate resistivity without requiring large fluctuations in temperature. The use of traditional static charge gating such as in a transistor structure restricts the extent to which carrier density can be altered, and ultimately limits the magnitude change in electrical resistivity across the MIT [7]. On the other hand, hydrogenation has recently been used as a mechanism to control proton doping in perovskite nickelates, with the ability to isothermally modulate resistivity by roughly 8 orders of magnitude [1]. Later work performed on NdNiO$_3$ interrogated resistivity modulation far from the thermal MIT (in the metallic region at and above room temperature) in NdNiO$_3$ (NNO) due to hydrogen doping and found a similarly large contrast in resistivity tuning from electron localization [8,9].

In this work, we evaluate the thermal conductivity of NNO across the MIT for both pristine and hydrogenated thin films. The thermal conductivity of such films is critical for modeling the behavior and heat dissipation requirements of novel electronics and new computational architectures that take advantage of the MIT behavior described in literature, and large variations in total thermal conductivity could lead to a new avenue for the development of an active, solid-state thermal switch. Likewise, the modulation of thermal properties between pristine and hydrogenated states might serve as a secondary sensing attribute in devices like optical modulators. $NdNiO_3$ (NNO) thin films were deposited on (001) $LaAlO_3$ (LAO) single crystal substrate using magnetron sputtering at room temperature. The detailed deposition conditions are as follows: atmosphere: 40:10 sccm $Ar/O_2$ mixture; total pressure: 5 mTorr; target power: Ni (DC power, 70 W) +Nd (RF power, 170 W). The film thickness is ~ 50 nm. After deposition, all films were post-annealed at 500 °C for 24 h in open air to form the perovskite phase. Hydrogen doping into the NNO/LAO film was performed using an electrochemical method. The experiment was done in three-electrode configuration in an electrolyte of 0.1M NaCl. The graphite rod was used as a counter electrode and the Ag/AgCl (in 3.5 M KCl) was used as a reference electrode. The NNO/LAO film was employed as the working electrode. For a typical proton doping procedure, a voltage of -1.5 V (vs Ag/AgCl) was applied for 30 s.

---

[*] Author to whom correspondence should be addressed. Electronic mail: warzoha@usna.edu

We further separate the electronic ($\kappa_e$) and phononic ($\kappa_{ph}$) contributions to thermal conductivity in order to highlight their relative importance in driving thermal transport between pristine (NNO) and hydrogenated (H-NNO) films. This allows us to build a qualitative picture of phonon scattering mechanisms according to Matthiessen's Rule, described later. To do this, we isolate the electrical conductivity of the films through direct measurement using a four-point probe configuration. Four ~ 80 nm Au/5 nm Ti pads, stretching the width of the film, were deposited. Thin gold wires were silver-bonded to the pads and used to connect the sample to a resistance bridge. Annealing of the bonds at 120 °C for ~ 15 min was necessary to improve the contact resistance and the mechanical strength of the bond. Electric current was applied between the outer pads while the voltage drop between the inner leads was monitored. The temperature was varied using a Quantum Design PPMS and stabilized at each data point.

Here we measure the total thermal conductivity of each sample ( $\kappa_{tot} = \kappa_{ph} + \kappa_e$ ) using frequency-domain thermoreflectance (FDTR) as a function of temperature (from 110 K to 250 K). FDTR is an optical pump-probe thermoreflectance technique capable of measuring the thermal properties of nanoscale thin-films. Details of its basic operating principle can be found in ref [10], while details associated with our measurement system can be found in refs [11,12]. As with most other thermal characterization techniques, FDTR is (broadly speaking) a method in which a sample is heated and its temperature response monitored. Specifically, FDTR uses separate pump and probe beams to heat the sample and detect changes in temperature, respectively. The pump beam (488 nm, Coherent Genesis MX, 1 W) is passed through an electro-optic modulator (EOM) and focused down onto the sample surface to establish a sinusoidally modulated, frequency-dependent heating event. The sample in this work is coated with an ~ 80 nm Au/5 nm Ti transducer to absorb the pump's photonic energy and convert it to thermal energy ahead of the sample. The transducer doubles as a thermometer due to a well-characterized relationship between changes in reflectance and temperature. Thus, a separate probe beam is used to monitor changes in reflectance at the sample surface. Because the temperature lags behind the heating event imposed on the transducer surface, changes in the phase lag between the modulated heating event and the temperature response can be used to determine underlying thermal properties of the sample stack. In this work, we modulate the pump frequency between 500 Hz and 20 MHz in order to vary the thermal penetration depth into the sample and gain sensitivity to a variety of thermal properties across multiple underlying material layers. To regulate the sample's temperature, we utilize an optical cryostat (Janis VPF-100) with an accuracy of ± 0.01 K. Sample electrical conductivities are shown in Fig. 1 (a) alongside the total, electronic, and phononic thermal conductivities of the pristine and hydrogenated samples in Fig. 1 (b). Note that the electronic thermal conductivity is calculated according to the Wiedemann-Franz law [13]. The electronic contribution to thermal conductivity for the H-NNO sample is found to be negligible and is therefore not shown in Fig. 1 (b).

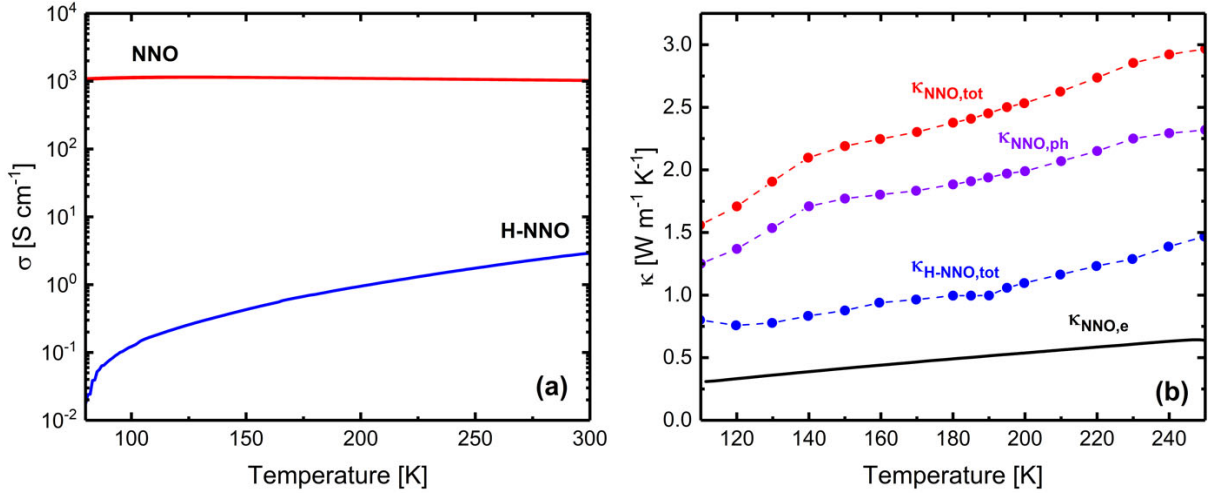

**Fig. 1:** (a) Temperature dependence of the electrical conductivity for NNO and H-NNO thin films and, (b) temperature dependence of the thermal conductivity for NNO and H-NNO thin-films with associated electronic and phononic contributions. Note that the total thermal conductivity of the H-NNO sample is nearly equivalent to the phononic contribution to its thermal conductivity given the low electrical conductivity of the H-NNO film.

The trend associated with the thermal conductivity of the NNO film is consistent with the thermal conductivity of bulk nickelate reported in [14]. Although few details are provided regarding the film's thickness, the results reported here are similar to the pristine NNO in [14], suggesting that the mean free path of heat energy carriers is much lower than the characteristic length of the film used in this work (~ 50 nm). While the MIT occurs at a different temperature (~ 165 K in this work, and 205

K in [14]), the contrast in thermal behavior between NNO and H-NNO obtained in this work should broadly apply to bulk system behavior.

A 3 to 5 orders of magnitude difference exists between the measured electrical conductivities of the NNO and H-NNO films. Interestingly, the disparity increases with a decrease in temperature below the MIT. This phenomenon is also reflected in the thermal conductivity at such temperatures, where a reduction in temperature generally corresponds to a larger discrepancy between the total thermal conductivity of NNO and H-NNO. Clearly, a reduction in the electrical contribution to the NNO film's thermal conductivity by introducing H atoms significantly reduces thermal performance. However, the full reduction in thermal conductivity cannot be attributed to the tuning of electrical conductivity alone. There is a nearly 50% decrease in the phononic contribution to thermal conductivity when H atoms are introduced into the crystal structure. Prior DFT simulations of NNO and 0.25H-NNO [15] reveal a nearly 2-fold reduction in the magnitude of the vibrational momentum across dominant low-frequency modes.

We use recent work by Li et al. [16] as a barometer for these results. In their work, the authors find a nearly 5-fold reduction in the thermal conductivity of $NdNiO_3$ upon protonation, and an even larger modulation in thermal conductivity between pronated and deprotonated $LaNiO_3$. These measurements were conducted at room temperature and agree well with the order of magnitude difference we see near room temperature in this work. Here, we additionally show that the thermal conductivity of the NNO film is substantially reduced below the MIT; however, the thermal conductivity ratio between the NNO and H-NNO films remains approximately constant as a function of temperature ($\kappa_{NNO}/\kappa_{H-NNO} \approx$ 2-3). While the largest disparity in thermal conductivity occurs above the MIT, the electronic contribution to thermal conductivity appears to have little to do with the degree to which the thermal conductivity can be modulated. This contrasts the suppositions in [16], where the authors found a more significant contribution by electrons. However, those analyses relied on a large modification to the Sommerfeld value of the Lorentz number when analyzing the electronic contribution to thermal conductivity (by a factor of more than 2). Although variation in the Lorentz number is known to occur for correlated material systems [17,18], this large an increase would be surprising for a complex oxide with relatively low electron-phonon coupling. Even if the Lorentz number is nearly double its accepted value of $L_0 = 2.44 \times 10^{-8}$ W • Ω • $K^{-2}$, the electronic contribution to thermal conductivity for our NNO films remains less than the phononic contribution. In such a case, the phononic thermal conductivity of the NNO film remains greater than that of the H-NNO film, suggesting that a significant fraction of the thermal conductivity reduction upon protonation is due to phonon scattering with the relatively large number of hydrogen atoms in the lattice.

Typically, low-level doping of light atoms in complex crystal structures does not result in meaningful changes to a host material's thermal conductivity. However, these hydrogenated films contain roughly 0.1 - 0.5 H-atom per unit cell, which introduces high mass-impurity scattering within the nickelate film. Given the ease with which these films can be hydrogenated, this film platform serves as an excellent candidate for active thermal conductivity switching, with up to a nearly 3-fold change in thermal conductivity at relevant device temperatures. Future work will leverage DFT simulations and analytical models to obtain the underlying mass-impurity scattering rates that govern this switching phenomenon.

**CONCLUSION**

In this work, we present the first evidence for modulation of thermal conductivity due to protonation in rare-earth nickelate films ($NdNiO_3$) across a broad range of temperatures. Frequency-domain thermoreflectance is used to determine the thermal conductivity of the films, and four-point electrical resistivity measurements are made to separate the electronic component of thermal conductivity from the phononic component. Results across a wide range of temperatures suggest that heavy protonation in such films completely suppresses the electronic contribution to thermal conductivity, but that the reduction in thermal conductivity does not appear to be the result of eliminating electrons as heat energy carriers alone. Instead, a high density of light atoms within the lattice structure likely results in significant mass-impurity scattering. Since hydrogen intercalation can be carefully controlled by choice of voltage and pulse width, the thermal conductivity can likely be tuned by electric fields akin to electrical properties. These results are critical for the design and optimization of thermal switching architectures that utilize rare earth nickelates and other emerging quantum materials and nearly degenerate semiconductors where extreme electron or hole doping is necessary to modulate carrier density.